\documentclass[%
reprint,
superscriptaddress,
showpacs,
showpacs,preprintnumbers,
amsmath,amssymb,
aps,
]{revtex4-1}

\usepackage[dvipdfmx]{graphicx}% Include figure files
\usepackage{dcolumn}% Align table columns on decimal point
\usepackage{bm}% bold math
\usepackage{multirow}
\usepackage[mathlines]{lineno}% Enable numbering of text and display math
\usepackage{comment}

\usepackage{xr}
\usepackage{color}
\begin{document}

\preprint{APS/123-QED}

\title{Magnetotransport studies of the Sb square-net compound LaAgSb$_2$ under high pressure and rotating magnetic fields}

\author{Kazuto~Akiba}
\email{akb@okayama-u.ac.jp}
\affiliation{
Graduate School of Natural Science and Technology, 
Okayama University, Okayama 700-8530, Japan
}

\author{Nobuaki~Umeshita}
\affiliation{
Graduate School of Natural Science and Technology, 
Okayama University, Okayama 700-8530, Japan
}

\author{Tatsuo~C.~Kobayashi}
\affiliation{
Graduate School of Natural Science and Technology, 
Okayama University, Okayama 700-8530, Japan
}

\date{\today}

\begin{abstract}
Square-net-layered materials have attracted attention as an extended research platform of Dirac fermions
and of exotic magneto-transport phenomena.
In this study, we investigated the magneto-transport properties of LaAgSb$_2$,
which has Sb-square-net layers and shows charge density wave (CDW) transitions at ambient pressure.
The application of pressure suppresses the CDWs, and above a pressure of 3.2 GPa, a normal metallic phase with no CDWs is realized.
By utilizing a mechanical rotator combined with a high-pressure cell,
we observed the angular dependence of the Shubnikov-de Haas (SdH) oscillation up to 3.5 GPa
and confirmed the notable two-dimensional nature of the Fermi surface.
In the normal metallic phase, we also observed a remarkable field-angular-dependent magnetoresistance (MR),
which exhibited a ``butterfly-like'' polar pattern.
To understand these results, we theoretically calculated the Fermi surface and conductivity tensor at the normal metallic phase.
We showed that the SdH frequency and Hall coefficient calculated based on the present Fermi surface model agree well with the experiment.
The transport properties in the normal metallic phase are mostly dominated by the anisotropic Dirac band,
which has the highest conductivity owing to linear energy dispersions.
We also proposed that momentum-dependent relaxation time plays an important role in the large transverse MR and negative longitudinal MR in the normal metallic phase,
which is experimentally supported by the considerable violation of Kohler's scaling rule.
Although quantitatively complete reproduction was not achieved,
the calculation showed that the elemental features of the butterfly MR could be reasonably explained as the geometrical effect of the Fermi surface.
\end{abstract}

\maketitle

%\tableofcontents
\section{Introduction}
\label{sec_intro}
A material class that possesses a two-dimensional square-net structure has attracted considerable attention in the field of condensed matter physics \cite{Klemenz_2019}.
In these materials, the bulk physical properties are governed by carriers with extremely high mobility (so-called Dirac fermions) characterized by linear energy dispersion, which causes various exotic magneto-transport phenomena.

Materials with the formula $MTP_{2}$ ($M$ is a rare-earth or alkali-earth metal, $T$ is a transition metal, and $P$ is a pnictogen element such as Bi and Sb) serve
versatile square-net compounds
\cite{Park_2011,Wang_2012,Masuda_2016,Li_2016,Liu_2016,Wang_2016,Liu_2017,Kealhofer_2018,Wang_2018,Sakai_2020,Liu_2021}.
One remarkable advantage of these materials is that systematic investigation is possible by choosing the appropriate $M$ and/or $T$ from a wide range of materials, including non-magnetic and magnetic elements.
The prominent quantum oscillation with the nontrivial Berry phase
\cite{Park_2011,Wang_2012,Masuda_2016,Li_2016,Liu_2016,Wang_2016,Liu_2017,Kealhofer_2018,Wang_2018,Sakai_2020,Liu_2021} and the quantum Hall effect in a bulk crystal \cite{Masuda_2016,Sakai_2020,Liu_2021}
have been regarded as hallmarks of Dirac fermions derived from the Bi- or Sb-square-net layers.

Another system being intensively studied is $M$X$Y$ ($M=$Zr, Hf \textit{etc}.,
$X=$Si, Ge, Sn,
and $Y=$S, Se, Te)
with a square net composed of $X$
\cite{Ali_2016,Lv_2016,Hu_2016,Kumar_2017,Hu_2017,Voerman_2019,Novak_2019,Chiu_2019,Shirer_2019}.
The intervention of the nontrivial electronic state in the experimentally observed large nonsaturating magnetoresistance (MR)
and high carrier mobility has attracted research interest in these materials.
Notably, the MR effect in these materials shows a strong field-angular dependence \cite{Ali_2016,Lv_2016,Hu_2016,Kumar_2017,Voerman_2019,Novak_2019,Chiu_2019,Shirer_2019},
which results in a butterfly-like shape on the polar plot of the resistivity.
Although active discussion on the origin of this ``butterfly MR'' has been performed,
whether it involves the nontrivial band structure or can be merely explained by a geometrical effect of the Fermi surface is currently unclear.

LaAgSb$_2$, the target of the present study, 
is classified into the former ``112-type'' material having Sb-square-net layers.
Reflecting the anisotropic crystal structure, the Fermi surface shows a remarkable two-dimensional nature \cite{Myers_1999b, Inada_2002},
which results in the unique physical properties of this material.
LaAgSb$_2$ exhibits successive two charge density wave (CDW) transitions at $T_{CDW1}\sim 210$ K and $T_{CDW2}\sim 190$ K at ambient pressure \cite{Myers_1999a, Song_2003}.
Previous X-ray diffraction studies have identified that CDW1 and CDW2 accompany lattice modulations
along the $a$- and $c$-directions, respectively \cite{Song_2003}.
The origin of these CDWs has been understood as a nesting of the corresponding part of the Fermi surface \cite{Bosak_2021}.
The fundamental physical properties, Fermi surface, and details of the CDW phase have been investigated by various research groups \cite{Arakane_2006, Budko_2008, Hase_2014, Chen_2017, Ruszala_2020}.
Interestingly, recent studies have proposed that LaAgSb$_2$ has a Dirac-like linear dispersion \cite{Wang_2012Oct, Shi_2016}, similar to the relevant square-net materials.
This linear dispersion is in the Fermi surface that is regarded to serve the nesting of CDW1.
Previous studies have suggested that the Dirac-like point in LaAgSb$_2$ is highly anisotropic \cite{Bosak_2021}:
it has a steep linear dispersion along high symmetry lines, whereas it is quadratic along the normal directions.
A similar highly anisotropic Dirac dispersion has been reported for SrMnBi$_2$ and CaMnBi$_2$ \cite{Park_2011, Lee_2013, Feng_2014}.

The authors recently reported the magneto-transport properties of LaAgSb$_2$ under high pressure \cite{Akiba_2021},
in which CDW1 and CDW2 were suppressed by pressure and disappeared at the critical pressures $P_{CDW1}\sim 3.2$ GPa and $P_{CDW2}\sim 1.7$ GPa, respectively.
We observed large positive transverse MR in the range of 1000--4000 \%
and characteristic negative longitudinal MR.
In addition, we observed a distinct change in the Shubnikov-de Haas (SdH) oscillation pattern at $P_{CDW1}$ and $P_{CDW2}$,
highlighting the drastic change of the Fermi surface.
The collapse of CDW1 causes an abrupt increase in the conductivity components $\sigma_{xx}$ and $\sigma_{xy}$,
suggesting a highly conductive electronic state above $P_{CDW1}$.
We can expect that above $P_{CDW1}$, the Fermi surfaces hidden by the CDW gap can fully contribute to the transport properties.
Thus, novel transport phenomena caused by the linear energy dispersion should be an intriguing issue.
In the previous report, the magnetic-field direction was fixed to the out-of-plane [001] direction,
and thus, details of the Fermi surface geometry under pressure have remained unexplored.
Various Fermi surface models have been proposed \cite{Myers_1999b,Inada_2002,Hase_2014,Ruszala_2020,Bosak_2021},
although there is no commonly recognized picture at present.

In the present study, we investigated the magneto-transport properties of LaAgSb$_2$
with combined experimental and computational methods.
We revealed the angular dependence of MR under a high pressure of up to 3.5 GPa utilizing a uniaxial mechanical rotator combined with a high pressure cell.
The angular dependence of the SdH frequencies observed in the high pressure phases supports the cylindrical geometry of the Fermi surface.
Furthermore, we observed a clear butterfly MR only above $P_{CDW1}$, which is similar to that observed in $MXY$ systems.
To discuss the experimental results, we calculated the Fermi surface under pressure
and the conductivity tensor based on the first-principles band structure.
Above $P_{CDW1}$, the largest contribution on the electrical conductivity was caused by the anisotropic Dirac band.
This dominance can be explained by the large Fermi velocity distributed over a wide range of the surface,
which is caused by the linear dispersion.
The experimentally obtained Hall coefficient
and the SdH frequency observed above $P_{CDW1}$ were reasonably explained by the present Fermi surface model.
We also showed that fundamental features of the butterfly MR can be qualitatively explained
by the geometrical effect of the Fermi surface.
From the quantitative viewpoint, however,
the high-field transverse MR and negative longitudinal MR were not reproduced by calculation.
We proposed that this inconsistency between the experimental and calculated results is caused by the momentum-dependent relaxation time,
which is not considered in the present calculation.
This assumption is supported experimentally by the violation of Kohler's scaling rule above $P_{CDW1}$.

\section{Experimental methods}
\label{sec_exp}
Single crystals of LaAgSb$_2$ were synthesized by the Sb self-flux method \cite{Myers_1999a}.
Detailed properties of the utilized samples are described in the previous report \cite{Akiba_2021}.

The resistivity under high pressure was measured
using an indenter-type pressure cell ($P < 3.5$ GPa) \cite{Kobayashi_2007}.
Daphne oil 7474 \cite{Murata_2008} was used as a pressure medium.
The pressure in the sample space was determined based on the superconducting transition temperature of Pb set near the sample.

The resistivity measurements under the condition of $B < 8$ T and $T > 1.6$ K were performed using a superconducting magnet and variable-temperature insert (Oxford Instruments).
Resistivity measurements under low temperature (0.1 K $< T <$ 1.6 K) and
high magnetic field (B $<$ 12 T) were performed using a combined system of a home-made
dilution refrigerator and superconducting magnet (Oxford Instruments).
LR-700 (Linear Research) or model 370 (Lake Shore Cryotronics, Inc.) AC resistance bridges were utilized for resistivity measurements.
We adopted the standard four-terminal method, and electrical contacts were formed by a silver paste (Dupont 4922N).

The angular dependence of the resistivity under pressure was measured using a home-made mechanical rotator, which can uniaxially rotate the indenter-type pressure cell in the variable-temperature insert.
The tilt angle of the pressure cell against the applied magnetic field was determined using a Hall sensor (HG-302C, Asahi Kasei Microdevices Corporation).

\section{Computational methods}
\label{sec_comp}
The band structure calculation based on the density-functional theory (DFT) was performed using the Quantum ESPRESSO package \cite{Giannozzi_2009, Giannozzi_2017} with full-relativistic ultrasoft pseudo potentials.
For calculations at ambient pressure (high-temperature phase without CDW),
we adopted the lattice constants $a=4.3941$ \AA, $c=10.868$ \AA, and atomic coordinates experimentally identified in our previous study \cite{Akiba_2021}.
For calculation under a pressure of 3.5 GPa, we adopted $a=4.3495$ \AA and $c=10.5598$ \AA,
which were obtained by linear extrapolation assuming the known lattice compressibility \cite{Budko_2006},
whereas the atomic coordinates were assumed to be identical with those at ambient pressure.
We used a cutoff of 75 Ry and 540 Ry for the plane wave expansion of
wave functions and charge density, respectively,
and a Monkhorst-Pack 6 $\times$ 6 $\times$ 6 $k$-point grid for the self-consistent calculation.

To investigate the details of the Fermi surface, we employed maximally localized Wannier functions (MLWF) method using Wannier90 \cite{Mostofi_2014}.
Based on the tight-binding Hamiltonian obtained by the MLWF method, we calculated the Fermi surface and electrical conductivity using WannierTools \cite{Wu_2017}.
For the Fermi surface calculation, we interpolated the full-relativistic DFT band structure with 44 Wannier functions (La-$d$ and Sb-$p$).
The Fermi surface and Fermi velocity
$\bm v_F=\hbar^{-1}\partial \epsilon(\bm k)/\partial \bm k|_{\epsilon = \epsilon_F}$
obtained using a dense $101\times 101\times 101$ $k$ mesh
were visualized
using FermiSurfer \cite{Kawamura_2019}.
Calculation of the conductivity tensor required relatively higher computational costs than the Fermi surface calculation.
Thus, we recalculated the scalar-relativistic DFT band structure with identical cutoffs and the $k$-point mesh mentioned above,
and reconstructed the tight-binding Hamiltonian with 22 Wannier functions for the conductivity tensor calculation.
As shown in the previous study \cite{Ruszala_2020} and Fig. S1 in Supplemental Material \cite{SM_URL}, the effect of the spin-orbit coupling caused a negligible effect on the shape of the Fermi surface,
and thus, no essential change in the conductivity tensor calculation is expected.
In Fig. S2 in Supplemental Material \cite{SM_URL}, we show a comparison between the DFT calculation and tight-binding model adopted for the conductivity calculation.
The tight-binding model reproduces the DFT calculation quite well.

The electrical conductivity tensor $\bm\sigma$ under a magnetic field was calculated based on the Boltzmann equation
within the relaxation-time approximation by WannierTools \cite{Zhang_2019}.
In the above framework, the conductivity tensor is represented by
\begin{equation}
\sigma_{ij}^{(n)}=\dfrac{e^2}{4\pi^3}\int d\bm k v_i^{(n)}(\bm k) \tau_n \bar{v}_j^{(n)}(\bm k)\left(-\dfrac{\partial f_{FD}}{\partial \epsilon}\right)_{\epsilon=\epsilon_n(\bm k)}.
\label{eq_CF}
\end{equation}
Here, $e$, $f_{FD}$, and $n$ represent the elemental charge, Fermi-Dirac distribution function, and band index, respectively.
$\tau_n$ represents the relaxation time of the $n$-th band, which is assumed to be independent of $\bm k$.
Because of the energy derivative of the Fermi-Dirac distribution function,
$\sigma_{ij}^{(n)}$ was determined by the states within the thermal energy width of $\sim k_B T$ near the Fermi level.
We set $T=10$ K to define the thermal energy width. 
$\bm v^{(n)}(\bm k)$ represents the velocity defined by the gradient of the energy in the reciprocal space as
\begin{equation}
\bm v^{(n)}(\bm k)=\dfrac{1}{\hbar}\dfrac{\partial \epsilon_n(\bm k)}{\partial \bm k}.
\end{equation}
$\bar{\bm v}^{(n)}(\bm k)$ represents the weighted average of velocity over the orbit,
which is defined as
\begin{equation}
\bar{\bm v}^{(n)}(\bm k)=\int_{-\infty}^{0}\dfrac{dt}{\tau_n}e^{t/\tau_n}\bm v^{(n)}[\bm k(t)].
\label{eq_wv}
\end{equation}
The historical motion of $\bm k(t)$ under a magnetic field $\bm B$ was obtained by the equation of motion
\begin{equation}
\dfrac{d\bm k(t)}{dt}=-\dfrac{e}{\hbar}\bm v^{(n)}[\bm k(t)]\times \bm B,
\end{equation}
where $\bm k(t=0)=\bm k$.
We adopted a $101\times 101\times 101$ $\bm k$ mesh for the calculation under fixed $\bm B$ (Fig. \ref{fig7})
and a $81\times 81\times 81$ $\bm k$ mesh for the field-angular dependence calculation (Fig. \ref{fig9}).

\section{Results}
\subsection{In-plane magnetoresistance}
First, we show the experimental results obtained under high pressure and rotating magnetic fields.
In this study, the magnetic field was tilted from the [001] axis to the normal directions to discuss the dimensionality of the Fermi surface.
$\theta$ is defined as a field angle measured from the [001] axis.
We utilized the term ``CDW1+2 phase'' ($0<P<P_{CDW2}$, CDW1 and CDW2 coexists), ``CDW1 phase'' ($P_{CDW2}<P<P_{CDW1}$, only CDW1 survives), and ``normal metallic phase'' ($P>P_{CDW1}$, no CDW exists) to specify the electronic state.
As mentioned in the previous report \cite{Akiba_2021}, the amplitude of the SdH oscillation in the CDW1 phase does not obey the conventional Lifshitz-Kosevich formula and takes the maximum at a relatively higher temperature.
Thus, most data were taken at 20 K in the CDW1 phase, unless otherwise specified.

\begin{figure}[]
\centering
\includegraphics[]{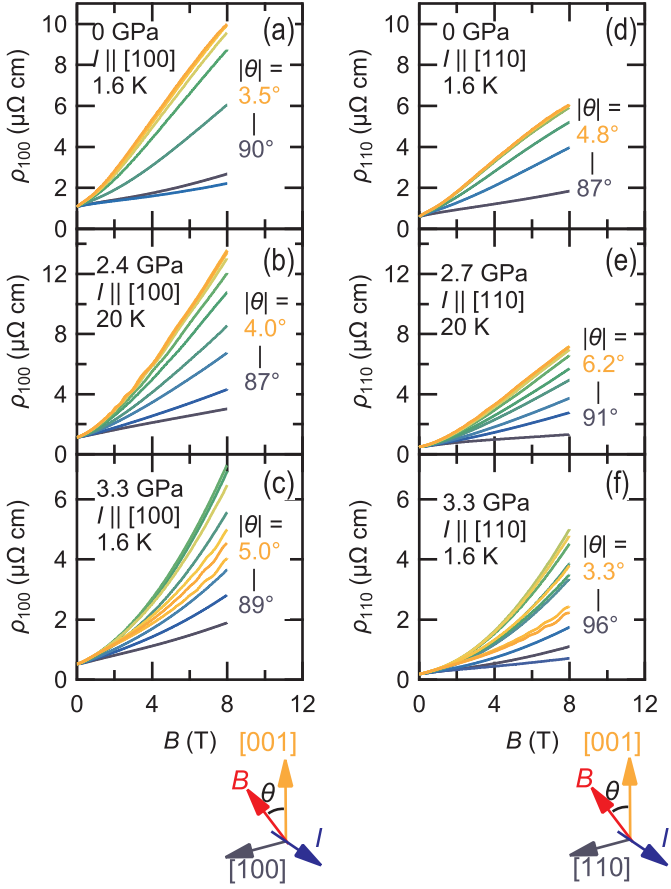}
\caption{
Magnetoresistance $\rho_{100}$ with current along the [100] direction at (a) 0 GPa, (b) 2.4 GPa, and (c) 3.3 GPa.
Magnetoresistance $\rho_{110}$ with current along the [110] direction at (d) 0 GPa, (e) 2.7 GPa, and (f) 3.3 GPa.
$\theta$ is defined by the angle between $B$ and [001] as shown in the schematic drawings.
The increment of $\theta$ from $0^\circ$ to $90^\circ$ corresponds to the color change from light orange to dark blue.
\label{fig1}}
\end{figure}

In Fig. \ref{fig1}, we show the field-angular dependence of the in-plane magnetoresistance ($\rho_{100}$ and $\rho_{110}$),
where electrical current ($I$) flows along the [100] and [110] directions.
In these measurements, the magnetic field ($B$) was rotated in the plane perpendicular to $I$.
At all phases, remarkable changes in the magnetoresistance were observed depending on the field direction.
We also observed an angular dependence of the SdH oscillation, which will be discussed later.
In the CDW1+2 and CDW1 phases, the angular dependence of the MR is rather simple:
it shows the largest positive MR effect in the case of $B\parallel [001]$ (light orange traces in Fig. \ref{fig1}),
and then gradually decreases as $B$ approaches the in-plane [100] or [110] directions (dark blue traces in Fig. \ref{fig1}).
In the normal metallic phase, however, the MR effect takes the maxima at intermediate angles as $\theta$ increases,
and then, takes the minimum around $\theta =90^\circ$.

\begin{figure}[]
\centering
\includegraphics[]{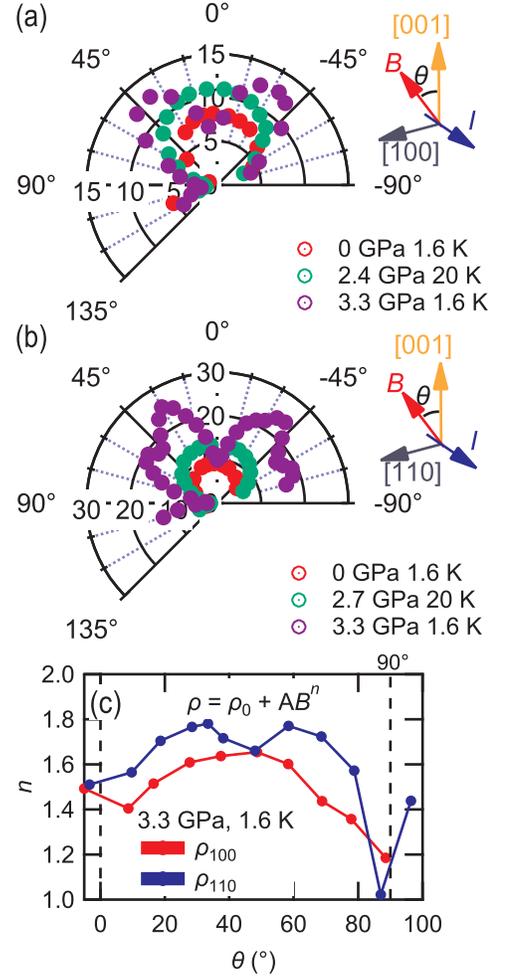}
\caption{
Polar plot of (a) $\Delta \rho_{100}/\rho$ and (b) $\Delta \rho_{110}/\rho$ at 8 T
in CDW1+2, CDW1, and the normal metallic phases.
(c) Exponent $n$ of $\Delta \rho_{100, 110}/\rho$ as a function of $\theta$.
Here, we assumed $\Delta \rho_{100, 110}/\rho \propto B^n$.
\label{fig2}}
\end{figure}

This change in the MR is clearly visible in the polar plots shown in Fig. \ref{fig2}(a) and (b).
In Fig. \ref{fig2}(a) and (b), the radius data represent the magnetoresistance at 8 T normalized by the value at 0 T
defined by $\Delta \rho_{100, 110}/\rho \equiv \rho_{100, 110}(8 T)/\rho_{100, 110}(0 T)-1$.
In the CDW1+2 and CDW1 phases, it shows simple and almost identical two-fold symmetry in both current configurations, and
the polar spectra are insensitive to the current direction.

In the normal metallic phase, in contrast,
polar spectra show a more complicated structure and obviously depend on the current direction.
The polar patterns show butterfly-like shapes in both cases, which are similar to those observed in the other square-net systems \cite{Ali_2016,Lv_2016,Hu_2016,Kumar_2017,Voerman_2019,Novak_2019,Chiu_2019,Shirer_2019}.
In case of $I \parallel$ [100], the magnetoresistance takes broad local maxima around $\pm 45^{\circ}$.
In case of $I \parallel$ [110], we can recognize the distinctive local maxima around $\pm 30^{\circ}$ and $\pm 60^{\circ}$.
In contrast to CDW1+2 and CDW1 phases, $\theta = 0^{\circ}$ changes into the local minima in both configurations.

In Fig. \ref{fig2}(c),
we summarize the $\theta$ dependence of $n$,
assuming the field dependence to be $\Delta \rho_{100, 110}/\rho \propto B^n$.
Around the local maxima of $\Delta \rho_{100, 110}/\rho$,
$n$ is slightly enhanced in both cases.
At $\theta=90^{\circ}$, in contrast, 
$n$ considerably decreases, and even shows almost linear field dependence.
At any $\theta$, $n$ is significantly smaller than conventional $n=2$.

The drastic change in the magneto-transport properties at $P_{CDW1}$ mentioned above is evidently
accompanied by emergence of a Fermi surface,
which has been gapped out by CDW1.
Because the angular dependence of the magnetoresistance roughly reflects the symmetry of the Fermi surface
cut perpendicular to the current direction \cite{Zhang_2019},
the two-fold symmetry is expected to be trivial considering the tetragonal crystal structure.
Additional structures observed only in the normal metallic phase are expected to contain information of the Fermi surface geometry.
This behavior will again be discussed in later sections.

\subsection{Out-of-plane magnetoresistance}

\begin{figure}[]
\centering
\includegraphics[]{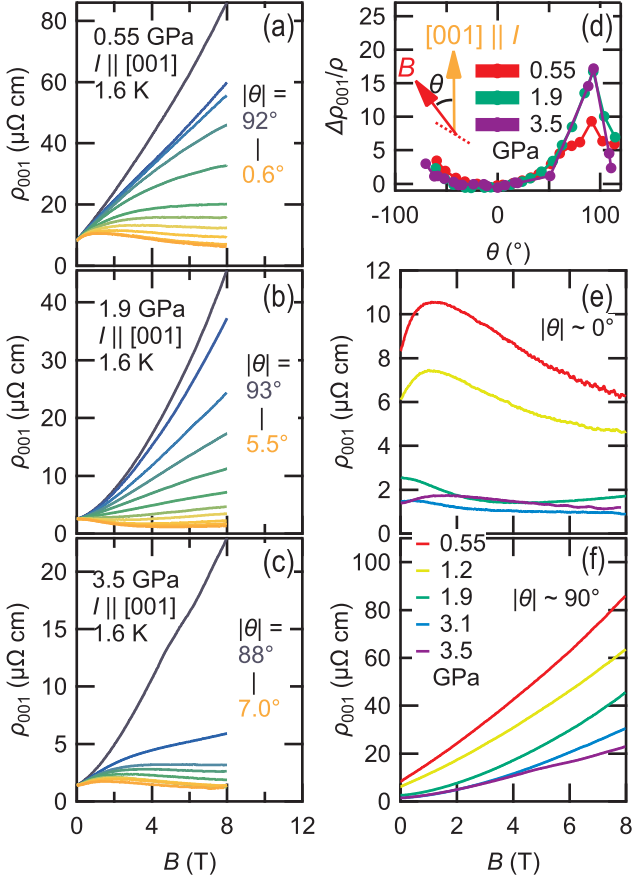}
\caption{
Magnetoresistance $\rho_{001}$ with current along the [001] direction at (a) 0.55 GPa, (b) 1.9 GPa, and (c) 3.5 GPa.
(d) Angular dependence of $\Delta \rho_{001}/\rho$ at 8 T.
Magnetic-field dependence of $\rho_{001}$ at several pressures around (e) $|\theta|=0^{\circ}$ and (f) $|\theta|=90^{\circ}$.
\label{fig3}}
\end{figure}

Next, we focus on the angular dependence of the out-of-plane magnetoresistance ($\rho_{001}$).
The in-plane crystal orientation has not been identified for this sample.
$\theta=0^{\circ}$ and $\theta=90^{\circ}$ correspond the longitudinal ($I\parallel B$) and transverse ($I \perp B$) configurations, respectively.

Figure \ref{fig3}(a), (b), and (c) show $\rho_{001}$ under various field directions in CDW1+2, CDW1, and normal metallic phases, respectively.
The angular dependence of $\Delta \rho_{001}/\rho$ at 8 T is summarized in Fig. \ref{fig3}(d).
Common to all phases, the magnetoresistance is considerably small
and even decreases as $B$ increases when $B\parallel [001]$ (light orange),
whereas it is robustly enhanced around $\theta=90^{\circ}$ (dark blue).
This behavior introduces very sharp peaks at around $\theta=90^{\circ}$ in Fig. \ref{fig3}(d).
$\Delta \rho_{001}/\rho$ at $\theta=90^{\circ}$ is smaller in the CDW1+2 phase and becomes approximately two times larger in the CDW1 and normal metallic phases.
The value of $\Delta \rho_{001}/\rho$ is not substantially different between the CDW1 and normal metallic phases.
Figure \ref{fig3}(e) and (f) show $\rho_{001}$ at representative pressures around $|\theta|\sim0^{\circ}$ and $90^{\circ}$,
respectively.
For $|\theta|\sim0^{\circ}$, the characteristic negative magnetoresistance is well consistent with our previous result \cite{Akiba_2021}.
For $|\theta|\sim90^{\circ}$, in contrast, negative contribution is completely absent, and $\rho_{001}$ increases with the application of $B$ without saturation.

\subsection{Angular dependence of the SdH oscillation}

\begin{figure}[]
\centering
\includegraphics[]{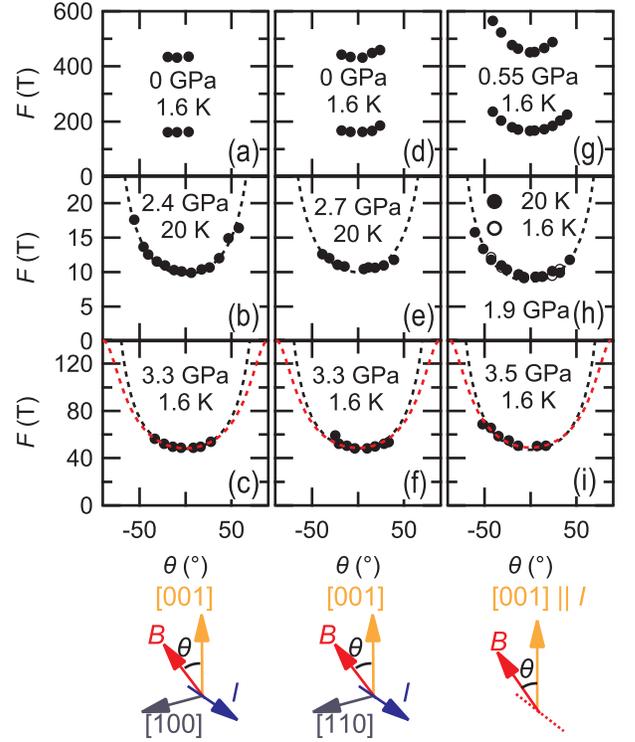}
\caption{
Angular dependence of the SdH frequency at each phase with current along the (a-c) [100], (d-f) [110], and (g-i) [001] directions.
The first, second, and third rows correspond to the CDW1+2, CDW1, and normal metallic phases, respectively.
The black broken lines represent the $1/\cos\theta$-type angular dependence,
and the red broken lines represent the ellipsoidal angular dependence with anisotropic parameter $e=2.9$ (see main text for detail).
\label{fig4}}
\end{figure}

Then, we move onto the angular dependence of the SdH oscillations superimposed on $\rho_{100}$, $\rho_{110}$, and $\rho_{001}$.
Figure \ref{fig4} shows the angular dependence of the SdH frequency ($F$).
The current and field configurations are illustrated at the bottom of each column.
The corresponding raw data are shown in Fig. S3 in Supplemental Material \cite{SM_URL}.

At ambient pressure (the first row in Fig. \ref{fig4}), we observed two frequencies, $F\sim160$ T and $\sim440$ T, at $\theta=0^{\circ}$,
which correspond to the $\beta$ and $\gamma$ branches, respectively, in the previous studies \cite{Myers_1999b, Inada_2002, Budko_2008}.
Each frequencies increases as the magnetic field is tilted from [001] axis,
which is consistent with the previous quantum oscillation measurements at ambient pressure \cite{Myers_1999b, Inada_2002}.

In the CDW1 phase (the second row in Fig. \ref{fig4}), we detected a single frequency of 10 T at $\theta=0^{\circ}$,
which is consistent with our previous report \cite{Akiba_2021}.
The angular dependence of this frequency well obeys the $1/\cos\theta$-scaling over the wide $\theta$ range
(up to $\theta=60^{\circ}$),
indicating the cylindrical geometry.
This indicates the existence of a considerably narrow cylindrical Fermi surface in the CDW1 phase.
We also confirmed that the $\theta$ dependence of $F$ is almost identical at the lower temperature of 1.6 K,
as shown by the open markers in Fig. \ref{fig4}(h).

\begin{figure}[]
\centering
\includegraphics[]{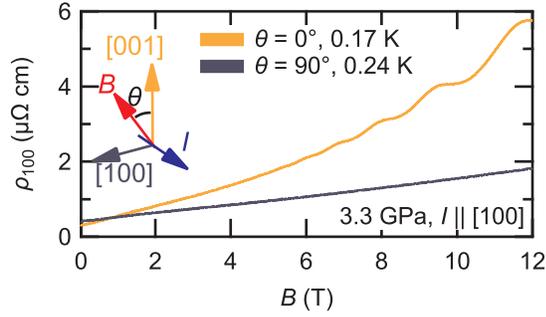}
\caption{
Magnetoresistance $\rho_{100}$ at $\theta=0^{\circ}$ and $90^{\circ}$ below 1 K
measured using dilution refrigerator.
\label{fig5}}
\end{figure}

In the normal metallic phase (the third row in Fig. \ref{fig4}),
the SdH oscillation has a single component in all configurations, whose frequency $F\sim 50$ T at $\theta=0^{\circ}$ is consistent with that of the $\omega$ branch identified in the previous study \cite{Akiba_2021}.
The obtained data are relatively poor compared to those in the CDW1 phase because of the significant decay of the oscillation amplitude as $\theta$ increases.
In the case in which the cross section comes from an elongated ellipsoid characterized by shorter ($a$) and longer ($b$) axes,
the $\theta$ dependence is proportional to
$1/\sqrt{\cos^2\theta +e^{-2}\sin^2\theta}$
where $e=b/a$.
The above function shows similar $\theta$ dependences at a small $\theta$.
We show the expected $\theta$ dependence when $e=2.9$ in Fig. \ref{fig4}(c), (f), and (i) by red broken lines.
For the $\omega$ branch, it is difficult to discuss the geometry only from these data.
Thus, we focus on the low-temperature MR measured at 0.24 K for $\theta=90^{\circ}$,
which is shown in Fig. \ref{fig5}.
When the SdH oscillation comes from a cylindrical geometry,
the orbit is open when $\theta=90^{\circ}$, and thus, no SdH oscillation is expected.
In case of ellipsoidal geometry, in contrast, there exists an extremum cross section at $\theta=90^{\circ}$,
and thus, it is reasonable to expect an SdH oscillation under sufficiently small thermal damping.
As seen in Fig. \ref{fig5},
oscillation is absent at the lowest temperature of the present study,
and hence, we regard the $\omega$ branch resulting from a cylindrical geometry.

To summarize the SdH measurements, we conclude that Fermi surfaces detected in the high pressure phases seem to be cylindrical along the $z$ direction.
This fact indicates that the strong two-dimensional nature is maintained up to the normal metallic phase.

\section{Discussion}
\subsection{Band structure and the Fermi surface in the normal metallic phase}
Here, we discuss the Fermi surface in the normal metallic phase based on the band structure calculations.
Because accurate information on the crystal structure and atomic coordinates \textit{etc} within the CDWs is not available,
it is difficult to quantitatively discuss the electronic structure in the CDW1+2 and CDW1 phases.
In this study, therefore, we mainly focus on the normal metallic phase,
in which we can directly compare the calculation with experiments owing to the absence of the energy gap and band reconstruction by CDW.

\begin{figure}[]
\centering
\includegraphics[]{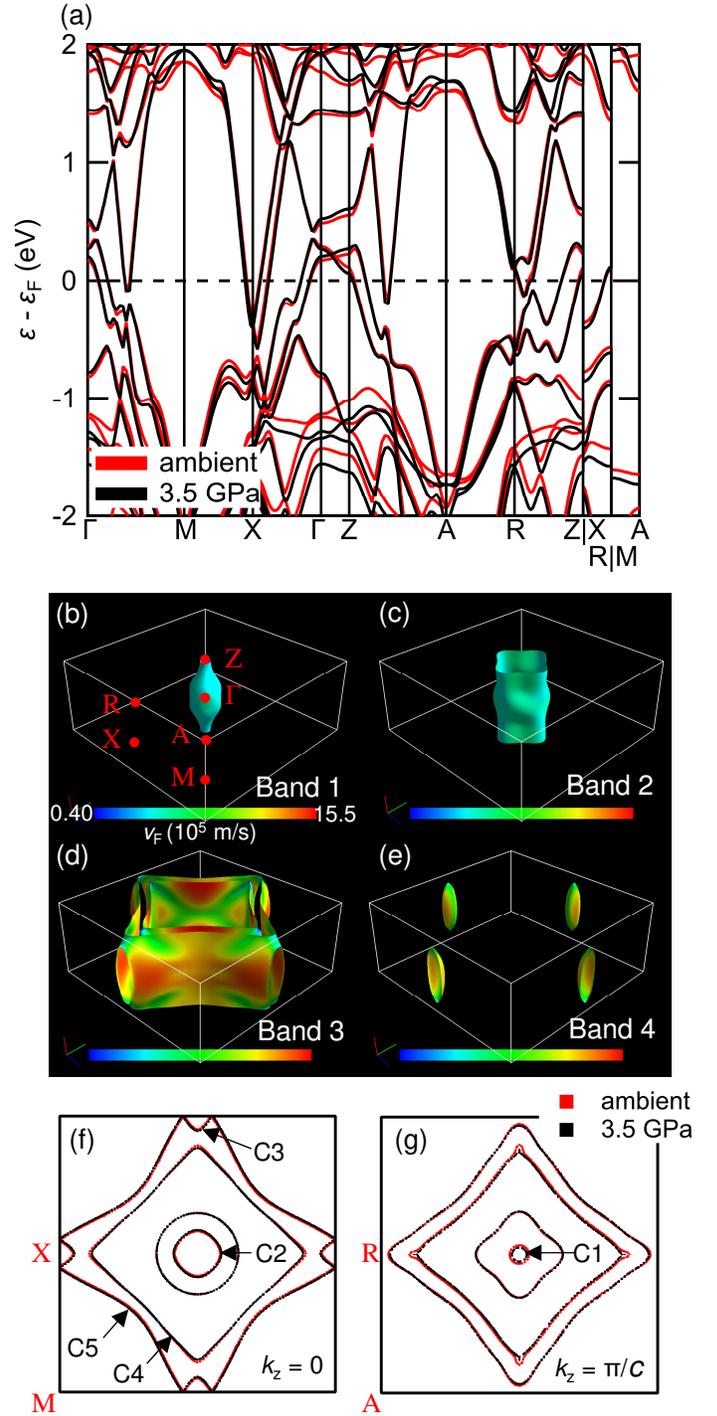}
\caption{
(a) Band structure at 3.5 GPa (black) obtained by full-relativistic DFT calculations.
The band structure at ambient pressure (red) is also overlaid for comparison.
(b-e) Fermi surfaces at 3.5 GPa originated from bands 1, 2, 3, and 4.
The color code represents the absolute value of the Fermi velocity $v_F$ from $0.40\times 10^5$ m/s (blue) to $15.5\times 10^5$ m/s (red).
Cross section of the Fermi surface cut by (f) $k_z=0$ and (g) $k_z=\pi/c$.
Black and red traces represent the cross-sections at 3.5 GPa and ambient pressure, respectively.
The outer black square represents the first Brillouin zone.
The size of the first Brillouin zone is equalized between the case of 3.5 GPa and ambient pressure. 
\label{fig6}}
\end{figure}

Figure \ref{fig6}(a) shows the band structure at 3.5 GPa (black) and ambient pressure (red).
We can see that at a pressure of 3.5 GPa, no remarkable change occurs near the Fermi level, such as the Lifshitz transition.
Figure \ref{fig6}(b-e) shows the Fermi surface of LaAgSb$_2$ at 3.5 GPa with the absolute value of the Fermi velocity $v_F$ shown by color code.
Hereafter, we refer to the bands constructing the surfaces shown in Fig. \ref{fig6} (b), (c), (d), and (e) as bands 1, 2, 3, and 4, respectively.
Bands 1 and 2 (3 and 4) are hole (electron) surfaces enclosing the higher- (lower-) energy region within the surface.
The hot spot of $v_F$ (e.g., around $X$ and on the way of $\Gamma$-$M$ and $Z$-$A$) corresponds to the point at which almost linearly dispersed bands cross $\epsilon_F$ in Fig. \ref{fig6}(a).
The maximum $v_F$ of $\sim 1.6\times10^6$ m/s is consistent with the previous calculation \cite{Ruszala_2020}
and comparable to those of SrMnBi$_2$ and graphene \cite{Park_2011}.
Compared with the previous calculation by Myers \textit{et al}. \cite{Myers_1999b} which is often referred in the discussion of quantum oscillation and nesting properties in this material, the shape of bands 2-4 is almost identical.
However, our calculation supports that band 1 is open along the $z$ direction,
which is a dice-like closed pocket in the calculation by Myers \textit{et al}.
The present Fermi surface geometry is well in accordance with a recent computational study
reported by Ruszala \textit{et al.} \cite{Ruszala_2020}.
In Tab. \ref{tab_SdH_calc}, we list the calculated cross section $S$ and the corresponding SdH frequency $F$ of several orbits at 3.5 GPa and ambient pressure.

We have shown in the experimental results that there exists a cylindrical Fermi surface with a small cross-section of $\sim$50 T in the normal metallic phase.
In our calculation, the orbit C1 around the $Z$ point in band 1 (Fig. \ref{fig6}(g)) can cause $1/\cos\theta$-type angular dependence, whose frequency is calculated to be 49.9 T at 3.5 GPa (Tab. \ref{tab_SdH_calc}).
This value shows reasonable agreement with the observed frequency.
Although orbit C2 in Fig. \ref{fig6}(f) can also cause $1/\cos\theta$-type dependence,
the frequency is calculated to be 486 T,
which is considerably larger than the observed frequency.
The orbit C3 in band 4 (Fig. \ref{fig6}(f)) shows comparable cross section with C1.
However,
the complete absence of the SdH shown in Fig. \ref{fig5} might be unlikely for the ellipsoidal shape.
The other cross-sections are considerably larger, and thus, we can exclude them from consideration.

According to the above results, we conclude that experimentally observed cross-section in the normal metallic phase originates from orbit C1, i.e., our result supports the cylindrical Fermi surface for band 1.
Conventionally, the CDW2 has been considered to be caused by the $k_z$-oriented nesting vector within band 1 \cite{Song_2003}, which was deduced from the dice-like shape of the Fermi surface.
Our conclusion suggests a careful reconsideration whether this nesting picture is valid in the case of the present Fermi surface geometry.

\begin{table}
\caption{\label{tab_SdH_calc}
Calculated cross section $S$ and SdH frequency $F$ of orbits C1-3 shown in Fig. \ref{fig6}(f) and (g) at 3.5 GPa.
The values in parenthesis were obtained from the calculation at ambient pressure.
}
\begin{ruledtabular}
\begin{tabular}{lrr}
orbit & $S$ (10$^{17}$ m$^{-2}$) & $F$ (T) \\
\hline
C1 & 4.76 (7.98) & 49.9 (83.6) \\
C2 & 46.4 (40.9) & 486 (429) \\
C3 & 14.8 (11.7) & 156 (123) 
\end{tabular}
\end{ruledtabular}
\end{table}

\subsection{Magneto-transport properties in the normal metallic phase in $B \parallel [001]$}
Here, we show the computational results of the conductivity and resistivity tensors in the normal metallic phase in $B \parallel [001]$ and compare them with the experimental results.
In the following calculation, the effect of Landau quantization is ignored.
This assumption is assumed to be reasonable
because the present situation is far from the quantum limit state.

\begin{figure*}[]
\centering
\includegraphics[]{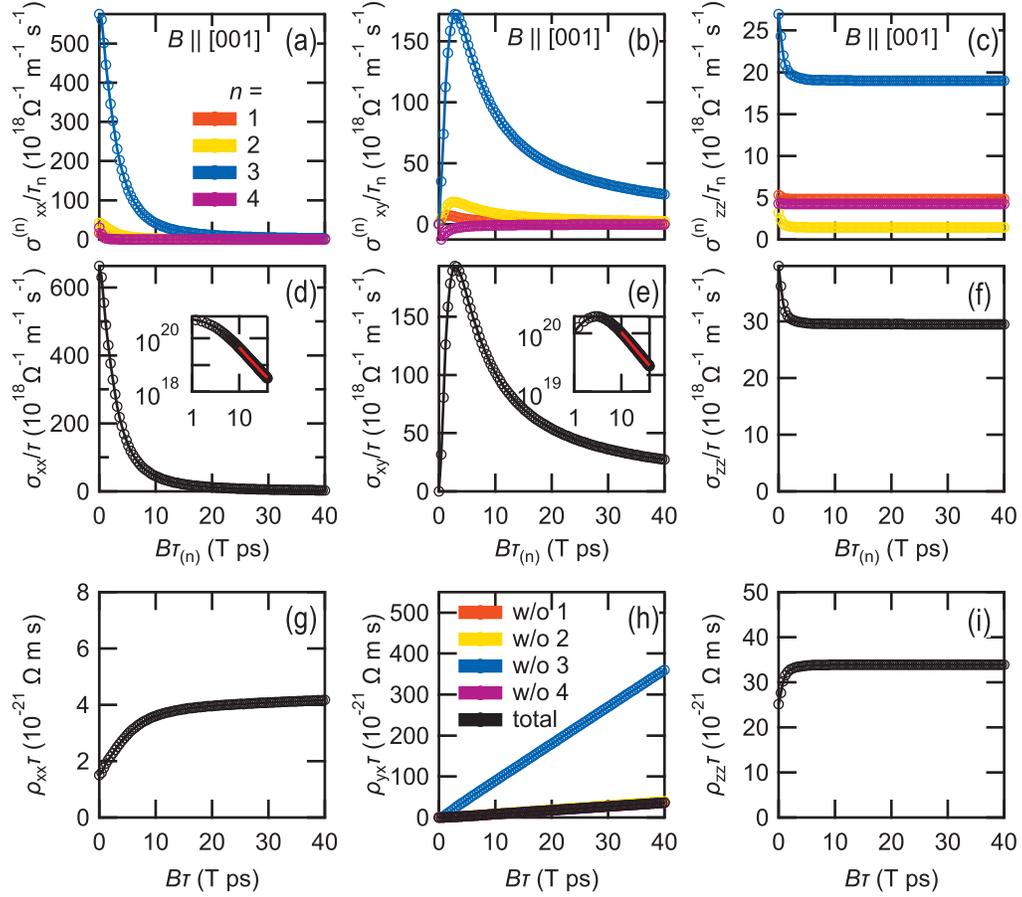}
\caption{
(a-f) Calculated conductivity components at 3.5 GPa with $B\parallel [001]$.
(a-c) are band-resolved $\sigma^{(n)}_{xx}/\tau_n$, $\sigma^{(n)}_{xy}/\tau_n$, and $\sigma^{(n)}_{zz}/\tau_n$ for $n=1$-4, and (d-f) are total conductivity $\sigma_{xx}/\tau$, $\sigma_{xy}/\tau$, and $\sigma_{zz}/\tau$,
obtained by the summation over $n$
assuming a $n$-independent relaxation time $\tau$.
The insets of (d) and (e) show the log--log plots of $\sigma_{xx,xy}/\tau$ as a function of $B\tau$.
Red lines indicate $n=-2$ for $\sigma_{xx}/\tau$ and $n=-1$ for $\sigma_{xy}/\tau$ assuming  $\sigma_{xx,xy}\propto B^{n}$.
(g-i) Calculated resistivity components $\rho_{xx}$, $\rho_{yx}$, and $\rho_{zz}$ at 3.5 GPa.
In (h), partial Hall resistivities for all possible cases are also shown (see main text for detail).
We can see that the absence of band 3 causes considerable changes in $\rho_{yx}$, whereas others bring a negligible effect.
\label{fig7}}
\end{figure*}

Figures \ref{fig7} (a-c) show the band-resolved conductivity components
$\sigma_{ij}^{(n)}/\tau_n$ at 3.5 GPa.
Here, the weight-averaged velocity in Eq. (\ref{eq_wv}) is arranged as a function of $B\tau_n$ as
\begin{equation}
\bar{\bm v}^{(n)}(\bm k, B\tau_n)=\int_{-\infty}^{0}\dfrac{d(Bt)}{B\tau_n}e^{Bt/B\tau_n}\bm v^{(n)}[\bm k(t)],
\end{equation}
and hence, $\sigma_{ij}^{(n)}/\tau_n$ is represented as a function of $B\tau_n$ as
\begin{equation}
\dfrac{\sigma_{ij}^{(n)}(B\tau_n)}{\tau_n}=\dfrac{e^2}{4\pi^3}\int d\bm k v_i^{(n)}(\bm k) \bar{v}_j^{(n)}(\bm k, B\tau_n)\left(-\dfrac{\partial f}{\partial \epsilon}\right)_{\epsilon=\epsilon_n(\bm k)}.
\label{eq_sigma_btau}
\end{equation}
The relaxation time in the $n$-th band
($\tau_n$) appears as an unknown parameter in the form of $\sigma_{ij}/\tau_n$ or $B\tau_n$ in the results shown in Fig. \ref{fig7}(a-c).
We recognize that in all components, the dominant contribution comes from band 3.
Based on Eq. (\ref{eq_sigma_btau}), the dominance of band 3 is explained
by the largest surface area and the distribution of high $v(\bm k)$ [Fig. \ref{fig6}(d)]
owing to its steep linear dispersion.
As shown in Fig. \ref{fig7}(b), bands 1 and 2 (band 4) lead to positive (negative) contributions on the Hall conductivity,
because these Fermi surfaces are hole (electron) surfaces and host only closed orbits when sliced by the (001) plane.
In contrast, Band 3 has a remarkable positive contribution,
although it is referred to in the conventional terminology as an electron surface.
The reason for this behavior is assumed to be the characteristic hollow-like geometry that enables the formation of hole-type closed orbits
when sliced by the (001) plane
(e.g., C4 and C5 in Fig. \ref{fig6}(f)).
The calculated $\sigma_{xy}/\tau$ indicates that LaAgSb$_2$ in the normal metallic phase
with $B\parallel [001]$ is
far from the compensation condition,
and the orbital character considerably inclines toward the hole-type.

By summing all band-resolved components, we obtain the total conductivity tensor divided by the relaxation time
($\sigma_{ij}/\tau=\sum_{n=1}^{4}\sigma_{ij}^{(n)}/\tau_n$),
whose non-zero components are shown in Fig. \ref{fig7}(d-f).
Here, we assume that the relaxation time is independent of the band index,
namely, $\tau_n=\tau$ for all four bands.
We can see in the insets of Fig. \ref{fig7}(d) and (e) that
$\sigma_{xx}\propto B^{-2}$ and $\sigma_{xy}\propto B^{-1}$ are satisfied,
which are generally known as the field dependence in the uncompensated case with no open orbit \cite{Chambers}.
By calculating the inverse matrix of $\sigma_{ij}/\tau$, we finally obtain the resistivity tensor $\rho_{ij}\tau$,
which can be compared with the experimental data.

In uncompensated metal with no open orbit,
the Hall resistivity no longer depends on the details of relaxation time
and can be determined only by the degree of compensation between electron-like and hole-like orbits in the high-field limit.
Thus,
the Hall resistivity is regarded to be a quantitative index of the Fermi surface geometry.
Here, we compare and discuss the experimental and theoretical Hall responses
in the normal metallic phase.

Fig. \ref{fig7}(h) shows the calculated Hall response $\rho_{yx}\tau$
as a function of $B\tau$.
The black curve is total $\rho_{yx}\tau$, calculated  by considering contributions from all bands.
It shows almost linear magnetic-field dependence up to $B\tau =40$ T ps.
This linear dependence is consistent with the experimental results \cite{Akiba_2021},
and thus, we can quantitatively compare the Hall coefficient $R_H$.
From the slope of the $\rho_{yx}\tau$-$B\tau$ plane,
we can theoretically evaluate $R_H=\rho_{yx}\tau/(B\tau)$,
which is a $\tau$-independent quantity.
We obtained $R_H=0.903\times 10^{-9}$ m$^3$/C from Fig. \ref{fig7}(h),
which shows reasonable agreement with the experimental value $R_H=1.16\times 10^{-9}$ m$^3$/C
at 3.4 GPa \cite{Akiba_2021}.
This agreement supports the validity of our Fermi surface model shown in Fig. \ref{fig6}.

We can also calculate the ``partial'' Hall resistivity, in which the contribution from a specific band is intentionally excluded when calculating $\bm\rho$.
This virtually describes the case in which a specific band is absent due to full CDW gap.
In Fig. \ref{fig7}(h), we show the partial Hall resistivity without the $n$-th band (w/o 1 to 4).
Obviously, the absence of bands 1, 2, and 4 hardly affects the slope of $\rho_{yx}\tau$,
and only hollow-like band 3 causes significant changes on $\rho_{yx}\tau$.
Previous experiments have shown that $\rho_{yx}$ at 9 T is reduced to 1/8
across $P_{CDW1}$ \cite{Akiba_2021}.
Such a large change can only be explained by the disappearance of band 3.
Thus, the above discussion confirms the dominant role of band 3
for the formation of CDW1.
However, we note that experimental $\rho_{yx}$ in the CDW1 phase shows unignorable non-linearity
in the weak-field region,
which is not reproduced by the calculated partial Hall resistivity.
In-depth knowledge regarding the Fermi surface within the CDW phase is indispensable to enter further quantitative aspects of the Hall response below $P_{CDW1}$.

Next, we move onto the transverse ($\rho_{xx}$) and longitudinal ($\rho_{zz}$) magnetoresistance.
Figure \ref{fig7}(g) and (i) shows $\rho_{xx}\tau$ and $\rho_{zz}\tau$ as a function of $B\tau$, respectively.
$\rho_{xx}\tau$ increases at the weak-field region
and then shows almost linear $B\tau$ dependence with a considerably small slope in $B\tau>10$ T ps.
$\rho_{zz}\tau$ also shows a sudden increase at a low $B\tau$ region and becomes almost flat above $B\tau>5$ T ps.
In the experimental result, in contrast,
both $\rho_{xx}$ and $\rho_{zz}$ show considerable field dependence:
$\rho_{xx}$ is proportional to $B^{\sim 1.5}$ as shown in Fig. \ref{fig2}(c),
and $\rho_{zz}$ shows characteristic negative longitudinal magnetoresistance as shown in Fig. \ref{fig3}(e).

Here, we discuss the reason for the mismatch between the experimental and calculated results.
The possible factor 
is the relaxation-time approximation adopted in this calculation.
We ignored the details of $\tau$, namely,
(1) $n$ dependence of $\tau$
and (2) $\bm k$ dependence of $\tau$.
As for (1), we can recognize, in the case of $\sigma_{zz}$, that
irrespective of how we set $\tau_n$ individually,
we cannot reproduce the decrease in $\rho_{zz}\sim 1/\sigma_{zz}$ observed in the experiment.
Thus, the factor (1) alone cannot explain the mismatch between the experiment and calculation.
To consider factor (2),
we have to introduce the $\bm k$-dependent relaxation time $\tau_n(\bm k)$
in the integrant in Eq. (\ref{eq_CF}).
Recent theoretical work on elemental metals showed that
there exists remarkable $\bm k$ dependence of $\tau$ even in such simple metals when the electron-phonon interaction is considered \cite{Mustafa_2016,Rittweger_2017}.
Because we do not know the precise representation of $\tau_n(\bm k)$,
we cannot discuss the effect of (2) quantitatively.
In the scope of the present study, factor (2) is not excluded from the possible cause.
Unlike the case of the Hall response,
the contribution of the relaxation time to the transverse and longitudinal magneto-transport does not vanish in the high-field limit,
and thus, they are affected by the details of the scattering process.

\begin{figure}[]
\centering
\includegraphics[]{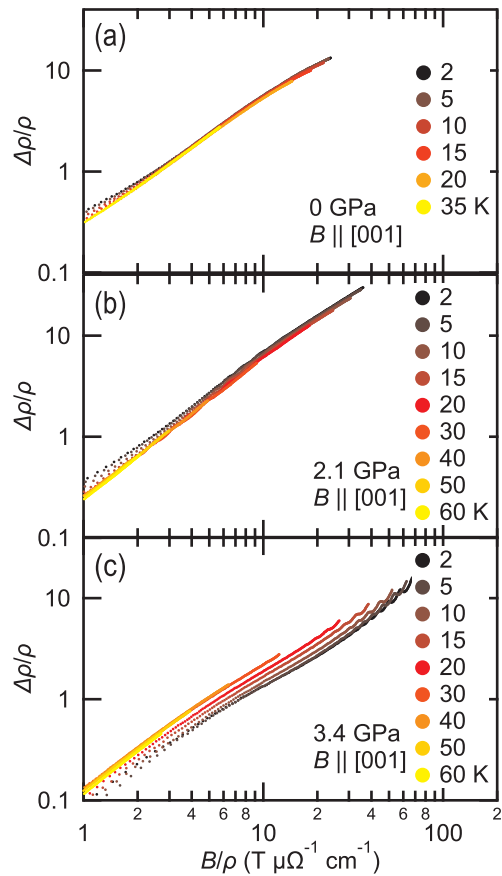}
\caption{
Kohler's plot at (a) ambient pressure, (b) 2.1 GPa, and (c) 3.4 GPa (see main text for details).
Here, the magnetic field is applied along the [001] direction and current along an in-plane direction.
\label{fig8}}
\end{figure}

The above consideration is also experimentally supported by the remarkable violation of the so called Kohler's rule in the normal metallic phase.
When the relaxation time is independent of $n$ and $\bm k$,
the transverse magnetoresistance should follow Kohler's scaling rule.
When we define $\Delta \rho/\rho \equiv [\rho(B, T)-\rho(0, T)]/\rho(0, T)$
where $\rho(B, T)$ represents the transverse magnetoresistance at magnetic field $B$ and temperature $T$,
Kohler's rule requires $\Delta \rho/\rho=f(B\tau)=F(B/\rho(0, T))$, where $f$ and $F$ represent arbitral functions, and we assume $\tau\propto 1/\rho(0, T)$.
To test the validity of Kohler's rule in the present case,
we construct a so called Kohler's plot for each phase, as shown in Fig. \ref{fig8}.
Here, the magnetic field is applied along [001] direction, and current along an in-plane direction.
In CDW1+2 and CDW1 phases, $\Delta \rho/\rho$ roughly falls into a universal curve as a function of $B/\rho(0, T)$
within the given temperatures,
indicating reasonable agreement with Kohler's rule.
In the normal metallic phase, however, 
the deviation from the Kohler's scaling becomes significant,
indicating that the picture of isotropic $\tau$ should be refined in the normal metallic phase.

\subsection{Angular dependence of the magneto-transport properties in the normal metallic phase}
Although we mentioned above that the conductivity calculation based on the conventional relaxation-time approximation is inadequate for a quantitative discussion on $\rho_{xx}$ and $\rho_{zz}$,
it is still insightful from the qualitative viewpoint to see whether the observed magneto-transport
properties can arise from the Fermi surface geometry.
Thus, we proceed with a comparison of the angular dependence of $\rho_{xx}$ and $\rho_{zz}$ between the experiment and calculation.
We performed similar calculations shown in Fig. \ref{fig7}(g) and (i)
for various $\theta$ with an interval of $2^{\circ}$ and constructed a polar plot.

\begin{figure}[]
\centering
\includegraphics[]{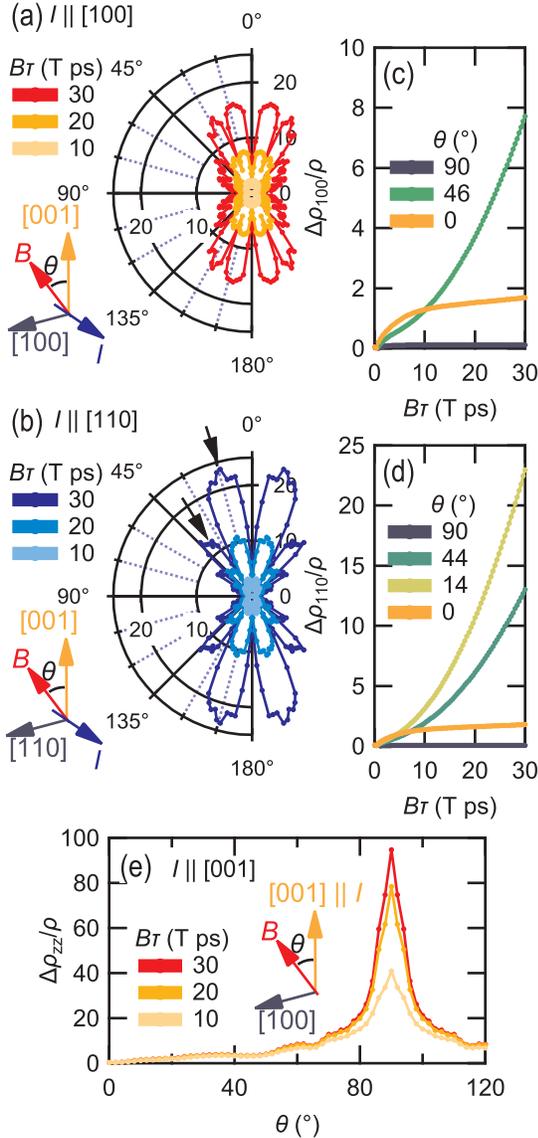}
\caption{Calculated polar plot at selected $B\tau$ (a, b)
and $B\tau$ dependence of magnetoresistance at representative $\theta$ (c, d)
for $I\parallel [100]$ and $I\parallel [110]$ cases.
(e) Calculated angular dependence of $\Delta \rho_{zz}/\rho$ at selected $B\tau$.
\label{fig9}}
\end{figure}

Figures \ref{fig9}(a) and (b) show the calculated polar plot of $\Delta \rho_{100}/\rho$ and $\Delta \rho_{110}/\rho$, respectively.
For several representative angles,
the $B\tau$ dependence of $\Delta \rho_{100}/\rho$ and $\Delta \rho_{110}/\rho$ is also shown in
Fig. \ref{fig9}(c) and (d).
We can see that remarkable field-angular dependence of the MR can arise from the Fermi surface geometry.
Further, we can recognize essential features observed in the experiment such as
the local minima at $\theta=0^{\circ}$ and 90$^{\circ}$ and a current-direction-dependent polar pattern.
A notable feature in the calculated polar plot is the very sharp decrease of the MR at specific field angles,
which results in a complex pattern.
As mentioned in Fig. S4 in Supplemental Material \cite{SM_URL}, these dips seem to arise in a certain manner,
which apparently resembles the Kajita--Yamaji oscillation observed in quasi-two-dimensional materials
\cite{Kajita_1989, Yamaji_1989}.
In the present stage, however, the specific mechanism of the reduction in resistance cannot be identified.
The changes in the MR are more rapid in the case of $I \parallel [100]$,
which occurs within $10^{\circ}$.
Focusing on the case of $I \parallel [110]$, the calculated polar pattern shows two distinct ``petals''
indicated by arrows in Fig. \ref{fig9}(b).
This feature qualitatively agrees with the experimental butterfly-like pattern shown in Fig. \ref{fig2}(b).

However, we should note several undeniable mismatches between the experimental and calculated results.
First, the theoretical polar pattern for $I \parallel [100]$ 
seems to have little resemblance with the experimental one [Fig. \ref{fig2}(a)].
As shown in Fig. S5 in Supplemental Material \cite{SM_URL},
the Fermi surface shift within $\pm50$ meV does not cause notable changes in the polar pattern.
We assume that coarse experimental data ($\sim 10^{\circ}$ interval)
and a slightly lower residual resistivity ratio (RRR) of the measured sample (RRR = 106 at 3.3 GPa for $I \parallel [100]$,
whereas RRR = 152 at 3.3 GPa for $I \parallel [110]$)
might mask the rapid change in the MR.
Second, the details of the polar pattern, e.g., the position of the maxima/minima for the $I \parallel [110]$ case and
the elongated shape of the calculated patterns disagree with the experimental results.
Although a slight difference in the Fermi surface curvature or shape can contribute to these mismatches,
we cannot specify the exact reason in the present study.
Because there can be seen several proposals for the possible Fermi surface \cite{Arakane_2006, Budko_2008, Hase_2014, Chen_2017, Ruszala_2020},
which are slightly different from each other,
the details of the geometry are open to further argument.
In addition, anisotropic $\tau$ revealed in the previous section
should also contribute to the above mismatches,
and hence, more accurate treatment for $\tau$ is indispensable for a quantitative discussion.

Finally, we show in Fig. \ref{fig9}(e) the calculated $\Delta \rho_{001}/\rho$ as a function of $\theta$.
In this calculation, $B$ directs [100] at $\theta=90^{\circ}$.
We can see a very sharp peak at around $\theta=90^{\circ}$,
and no remarkable structure is discernible.
This behavior qualitatively reproduces the experimental results.
The enhancement of MR around $\theta=90^{\circ}$ can be understood as the very small $v_z$ component
because most cross-sections are open along the [001] direction when the Fermi surface is sliced parallel to the [001].
From Eq. (\ref{eq_CF}), small $v_z$ results in small $\sigma_{zz}$, and thus, $\rho_{zz}$ becomes large.

To summarize, although complete agreement between the experiment and calculation is not achieved,
elemental features of the observed angular dependence of the MR can be qualitatively explained by the calculation.
Thus, we assume that the angular dependence of the magneto-transport properties observed in the normal metallic phase
is mainly derived from the geometrical effect of the Fermi surface.

\section{Conclusion}
We investigated the angular dependence of the magneto-transport properties and the Shubnikov-de Haas (SdH) oscillation of LaAgSb$_2$ under high pressure up to 3.5 GPa.
The angular dependence of the in-plane magnetoresistance (MR) in the normal metallic phase showed a butterfly-like polar pattern,
which has been reported in the relevant square-net system.
In high pressure phases, we identified cylindrical Fermi surfaces by the angular dependence of the SdH oscillation,
indicating that LaAgSb$_2$ holds a remarkable two-dimensional Fermi surface property up to 3.5 GPa.
Compared with the band structure and Fermi surface calculation at 3.5 GPa,
we conclude that the most inner cylindrical hole surface around the $\Gamma$ point is responsible for the $\omega$ branch observed in the normal metallic phase.
Our results suggest a careful reconsideration of the conventionally adopted nesting picture of the CDW2.
We also calculated the conductivity and resistivity tensors under a magnetic field
and showed reasonable agreement in terms of the Hall coefficient.
This provided strong support for the validity of the calculated Fermi surface and dominant contribution of hollow-like Fermi surfaces to the formation of CDW1.
In contrast, the large positive MR in $\rho_{xx}$ and negative MR in $\rho_{zz}$ were
not reproduced by the calculation.
A possible reason for this mismatch is assumed to be the momentum-dependent relaxation time,
which was experimentally supported by the violation of Kohler's rule.
The elemental features of the butterfly-like polar pattern were qualitatively reproduced by calculation,
and thus,
we considered that the remarkable anisotropy of the MR is basically caused by 
the geometrical effect of the Fermi surface.
From the quantitative viewpoint, however, there are undeniable mismatches between the experimental and calculated results.
To access these inconclusive issues, more accurate treatment of the relaxation time and further inspection of the details of the Fermi surface are necessary in future studies.

\begin{acknowledgments}
We thank H. Harima, J. Otsuki, and N. Hanasaki for variable comments and kind support.
This research was supported by JSPS KAKENHI Grant Number 19K14660.
\end{acknowledgments}

\bibliography{reference}% Produces the bibliography via BibTeX.

\clearpage

\end{document}